# Superconducting susceptibility signal captured in a record-wide pressure range

Shu Cai[1]*, Jinyu Zhao[1]*, Di Peng[1]*, Yazhou Zhou[2], Jing Guo[2], Nana Li[1], Jianbo Zhang[1], Yang Ding[1], Wenge Yang[1], Qiaoshi Zeng[1]†, Qi Wu[2], Tao Xiang[2], Ho-kwang Mao[1]† and Liling Sun[1,2]†

[1]*Center for High Pressure Science & Technology Advanced Research, Beijing 100094, China*
[2]*Institute of Physics, Chinese Academy of Sciences, Beijing 100190, China*

In recent years, the resistance signature of the high-temperature ($T_c$) superconductivity above 250 K in highly-compressed hydrides ( >100 GPa) has garnered significant attention within the condensed matter physics community. This has sparked renewed optimism for achieving superconductivity under room-temperature conditions. However, the superconducting diamagnetism, another crucial property for confirming the superconductivity, has yet to be conclusively observed. The primary challenge arises from the weak diamagnetic signals detected from the small samples compressed in diamond anvil cells. Therefore, the reported results of superconducting diamagnetism in hydrides have sparked intense debate, highlighting the urgent need for workable methodology to assess the validity of the experimental results. Here, we are the first to report the ultrahigh-pressure measurements of the superconducting diamagnetism on $Nb_{0.44}Ti_{0.56}$, a commercial superconducting alloy, in a record-wide pressure range from 5 GPa to 160 GPa. We present detailed results on factors such as sample size, the diamagnetic signal intensity, the signal-to-noise ratio and the superconducting transition temperature ($T_c$) across various pressures and different pressure transmitting media. These comprehensive results clearly demonstrate that this alloy is an ideal reference sample for evaluating superconductivity in compressed hydrides - validating the credibility of the experimental systems and superconducting diamagnetic results, as well as determining the nature of the superconductivity of the investigated sample. In addition, these results also provide the valuable benchmark for studying the pressure-induced superconductivity in other material families.

After the discovery of superconductivity in copper oxide and iron pnictide systems[1-3], scientists have been actively exploring new superconductors with higher superconducting transition temperatures ($T_c$). These efforts stand as the most significant issues for researchers in the fields of condensed matter physics and material sciences, as the materials with room temperature superconductivity will revolutionize our everyday life. In recent years, many metal hydrides exhibiting the zero-resistance signature of high-$T_c$ superconductivity with $T_c$ above 200 K under ultrahigh-pressure conditions have sparked fresh excitement in the field of high-$T_c$ superconductivity research[4-14]. Given that this resistance signature can be observed under ultrahigh-pressure conditions ( >100 GPa), the primary challenge in these investigations lies in assessing their superconducting diamagnetism, a crucial property for confirming the presence of superconductivity. It is known that the size of the sample subjected to more than 150 GPa is about 30-40 μm, resulting in a quite weak signal from the superconducting diamagnetism, as the intensity of the diamagnetic signal is directly proportional to the size of the sample. In the meantime, the diamagnetism obtained by various groups or through alternative methods is quite different, which stands as a main debating point among scientists[15-21].

To improve this situation and lead the community towards a greater consensus regarding the superconducting diamagnetism, there is an urgent need for an effective experimental method to assess the diamagnetic signal from the ultrahigh-pressure-induced superconducting transition. Recently, we proposed to use $Nb_{0.44}Ti_{0.56}$ alloy, a commercially available superconducting material with $T_c$ about 9.6 K at ambient pressure, as a standard reference sample to validate the reliability of the employed measurement systems and the results of the investigated samples with possible superconductivity under high pressure[22]. This approach can not only aid in confirming the intrinsic superconducting diamagnetism of the investigated sample but also the nature of superconductivity by comparing the intensity of the diamagnetic signal between the sample and the $Nb_{0.44}Ti_{0.56}$ alloy at a fixed pressure. Although the robust superconductivity of the $Nb_{0.44}Ti_{0.56}$ alloy against ultrahigh-pressure up to 261 GPa has been found by our previous measurement of persistent zero-resistance in the pressure range[23], its superconducting diamagnetic property has yet to be confirmed. In this study, we are the first to report the results of superconducting diamagnetism for $Nb_{0.44}Ti_{0.56}$ in a record-wide pressure range. The continuous change of its superconducting transition temperature, intensity of the diamagnetic signal, signal-to-noise ratio across various pressures up to 160 GPa and in different pressure transmitting media are provided.

Our high-pressure alternating current (*ac*) susceptibility measurements on superconducting $Nb_{0.44}Ti_{0.56}$ alloy were performed in a diamond anvil cell (DAC) fabricated from Cu-Be alloy, which is equipped with a gas membrane system (Fig. 1). Diamond anvils with 130 μm, 80 μm and 50 μm culet diameter (the flat area of the diamond anvil) were employed for independent measurements. To improve the sensibility of the measurement system, the modulated susceptibility [$\Delta\chi'=\chi'(B_m\neq0)-\chi'(B_m=0)$] technique was applied. This method generates a peak-like response characteristic of the superconducting transition, here $\Delta\chi$ represents the difference in susceptibility observed under a magnetic field ($B_m$) and zero magnetic field[18, 24-26].

Figure 2 presents the results of temperature versus modulated *ac* susceptibility ($\Delta\chi'$) measured from the $Nb_{0.44}Ti_{0.56}$ alloy for pressures ranging from 5 GPa to 160 GPa, which is a record-wide range for continuously capturing the diamagnetic signal of the superconducting transition, to the best of our knowledge. These results were obtained by conducting measurements in a DAC with an 80 μm-culet size and using NaCl as the pressure transmitting medium. It is found that the onset of superconducting diamagnetism starts at 9.6 K at the first pressure point of 5 GPa (Fig. 2a), as indicated by an arrow. In the pressure range of 5-49 GPa, the superconducting transition temperature ($T_c$) displays a continuous increase with pressure (Fig. 2a), in good agreement with the results from our resistance measurements[23] . In the ultra-high-pressure range of 116-160 GPa, the superconducting diamagnetic signal can still be observed (Fig. 2b). The superconducting transition temperature ($T_c$) detected by our *ac* susceptibility measurements displays a slight increase and then remains almost unchanged (Fig. 2b), also in good consistent with our resistance results reported previously[23]. Upon further increasing pressure above 160 GPa, as anticipated as the maximum pressure for the 80 μm culets, the diamond was broken at this pressure[27]. We performed identical measurements on the $Nb_{0.44}Ti_{0.56}$ alloy, employing smaller culets (about 50 μm) of the pressure anvils and only observed the diamagnetic single of the sample up to 28 GPa (see Supplementary Information-SI).

Given that the signal intensity of the superconducting diamagnetism is decreased apparently with increasing pressure below 49 GPa (Fig. 2a), we suppose that this may be attributed to a non-hydrostatic pressure environment. To clarify our conjecture, we conducted susceptibility measurements on $Nb_{0.44}Ti_{0.56}$ in a hydrostatic pressure environment, in which Helium was used as the pressure transmitting medium. As shown in Fig. 2c and 2d, the signal intensity of superconducting diamagnetism decreases slightly below 40 GPa, nevertheless, it remains relatively constant in the pressure range of 40-110 GPa. These results highlight the genuine impact of high-pressure environment on the superconducting diamagnetic signal.

Figure 3 illustrates the change in intensity of the superconducting diamagnetic signal with respect to pressure for $Nb_{0.44}Ti_{0.56}$. Notably, the signal intensity of the 40±2-μm sample compressed with 80 μm-culets in a NaCl medium exhibits a significant decrease in the pressure range of 5-49 GPa, from 114 mV at 5 GPa to 35 mV at 49 GPa. Similar phenomena have been also observed from the pressure-induced superconductivity in yttrium and $CaC_6$[28, 29]. As the pressure reaches high levels, specifically in the range of 116 -160 GPa, the signal intensity alters slightly, varying from 19 mV to 17 mV (Fig. 3a, 3c-3e), demonstrating the robust capability of diamagnetic signal agaist the ultra-high pressure. The signal intensity of the sample compressed with 50 μm-culets against pressure in a NaCl medium is also plotted in Fig. 3(a), revealing a similar trend. Given that the size (25±2 μm) of the sample compressed with 50 μm-culets is smaller than that (40±2 μm) of the sample compressed with 80 μm-culets, it is reasonable to observe that the ambient-pressure signal intensity of the former is lower than that of the latter within the low-pressure range.

The signal intensity of the superconducting diamagnetism versus pressure for the sample enclosed by Helium was also plotted in Fig. 3a. At 26 GPa, the signal intensity of the sample is about 77 mV, a value comparable to that obtained from the sample surrounded by a NaCl medium at a similar pressure level. As pressure is increased to 40 GPa, the signal intensity is about 66 mV and displays a slow decrease. Upon further increasing pressure from 42 to 110 GPa, the signal remains relatively stable, varying from 65 mV to 63 mV. This behavior differs from what observed from the sample compressed in a NaCl medium at the equivalent pressure range, where a rapid decrease in signal intensity occurs, demonstrating that the signal intensity is highly sensitive to the pressure condition associated with different pressure media. Therefore, it is advisable to emply Helium as the pressure transmissing medium in the measurements, when feasible.

To corroborate the $T_c$ value derived from our modulated *ac* susceptibility measurements on the $Nb_{0.44}Ti_{0.56}$ alloy, together with our direct-current (*dc*) susceptibility data (see SI), we established a pressure-$T_c$ phase diagram as shown in Fig. 4, and juxtaposed the results obtained from our resistance measurements[23]. It reveals that the pressure dependence of $T_{c\text{-}ac}$ and $T_{c\text{-}dc}$ values determined by our *ac* and *dc* susceptibility measurements and $T_{c\text{-}R}$ determined by our resistance measurements is well matched, reflecting the reliability of our experimental system and results. In this figure, $T_{c\text{-}ac}$ and $T_{c\text{-}dc}$ are the onset temperature of superconducting diamagnetism (Fig. 2), which is about the same value as that of 30% resistance drop ($T_{c\text{-}R}$).

In conclusion, we report the first set of superconducting diamagnetic results in a record-wide pressure range for the commercial $Nb_{0.44}Ti_{0.56}$ alloy, as summarized in Table 1. This summary includes information on the details of the diamond anvil and the sample size, $T_c$ values, intensities of superconducting diamagnetic signal, and signal-to-noise ratios at various pressures and in different pressure transmitting media (the use of Helium as the pressure transmissing medium is preferable). Based on our results of this study, we highlight the following two functions in the studies of ultrahigh-pressure-induced superconducting dimagnitism through the utilization "the reference sample method":

1. Verifying the dependability of the experimental system and the results of superconducting diamagnetism from the investigated sample.

   If the experimental system fails to detect the superconducting diamagnetism in the reference sample under pressure, it implies an inability to pick up any superconducting diamagnetic signal at this pressure. Consequently, the credibility of any diamagnetic results from the sample under investigation should be carefully analysized. It is advised to condact measurements on a sample that closely corresponds in size and thickness to the $Nb_{0.44}Ti_{0.56}$ sample.

2. Determining the nature of the superconductivity exhibited by the sample under investigation.

   If superconducting diamagnetism is detected, it is advisable to compare the signal intensity measured from the investigated sample (SI-sample) with that of the reference sample (SI-$Nb_{0.44}Ti_{0.56}$) at the corresponding pressure level. If the ratio of (SI-sample)/(SI-$Nb_{0.44}Ti_{0.56}$) approaches ~ 1, the investigated sample should possess bulk superconducting nature. While, if (SI-sample)/(SI-$Nb_{0.44}Ti_{0.56}$) is significantly deviates from 1, the sample should exhibit non-bulk superconducting behavior. In the latter case, the relative superconducting volume fraction ($f$) of the investigated sample can be roughly estimated using the formula: $f$ = [SI-sample ($P_i$)/(SI-$Nb_{0.44}Ti_{0.56}$ ($P_i$)]×100%, where $P_i$ represents a fixed pressure.

Moreover, the $Nb_{0.44}Ti_{0.56}$ alloy can also serve as a reference sample for the ultrahigh-pressure studies on superconducting diamagnetic property employed by alternative methods, such as flux traping and NV center methods[30-37], providing the researchers an available way to effectively validate their measurement system and the preduced results.

These authors with star (*) contributed equally to this work.

Correspondence and requests for materials should be addressed to Qiaoshi Zeng(zengqs@hpstar.ac.cn), Ho-kwang Mao (maohk@hpstar.ac.cn) and Liling Sun (liling.sun@hpstar.ac.cn or llsun@iphy.ac.cn).

## Acknowledgements

We thank Drs. Viktor Struzhkin, Dmitrii Semenok and Di Zhou for helpful discussions on this work. The work was supported by the National Key Research and Development Program of China (Grants No. 2021YFA1401800, 2022YFA1403900 and 2022YFA1402301), and NSF of China (Grants No., U2032214, 12122414 and 12274207). Partial work was supported by the Synergetic Extreme Condition User Facility (SECUF).

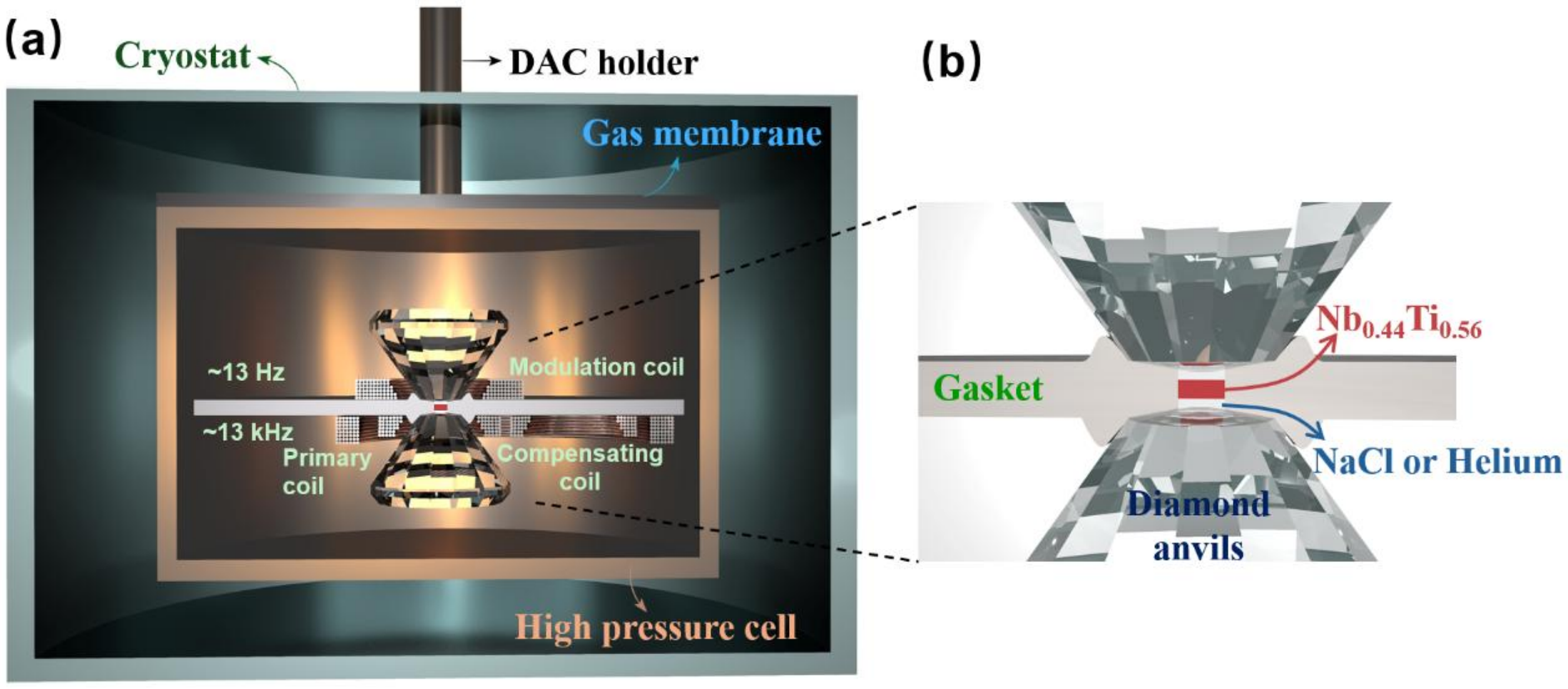

**Figure 1 Schematic description of experimental setup**. (a) The layout of the high-pressure cell (DAC) with a sample and coils in a cryostat. The sample is surrounded by a secondary coil (pickup coil) and a field-generating primary coil which is wound on the top of the secondary coil. For the primary coil, the alternating magnetic field was stimulated at 13 kHz. The exciting current and magnetic field are 5.8 mA and 5 mT, respectively. The modulated magnetic field is provided by the secondary coil that is stimulated at 13 Hz. The excited current and magnetic field are 27.6 mA and 10 mT, respectively. (b) Enlarged view for the assembling around the sample hole, including the arrangements for the non-magnetic gasket, $Nb_{0.44}Ti_{0.56}$ sample and pressure transmitting medium NaCl or Helium.

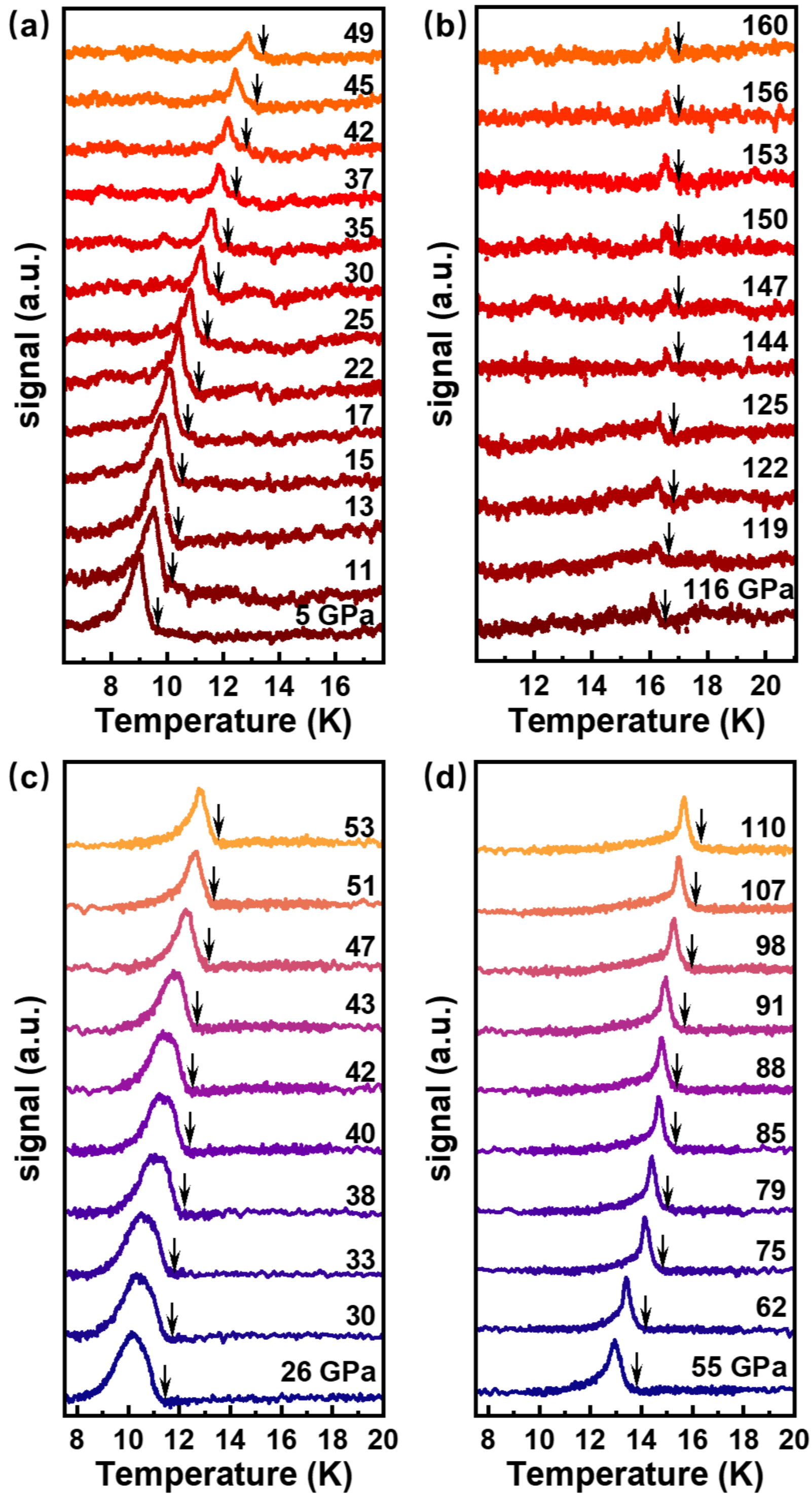

**Figure 2 The superconducting diamagnetic transition of $Nb_{0.44}Ti_{0.56}$ alloy detected by the modulated *ac* susceptibility measurement under high pressure with different pressure transmitting media.** (a) and (b) Temperature dependence of superconducting diamagnetic signal of the sample surrounded by NaCl medium for pressures ranging from 5 GPa to 160 GPa. (c) and (d) The superconducting diamagnetic signal versus temperature for the sample surrounded by Helium medium under pressure up to 110 GPa. The arrows indicate the superconducting transition temperatures.

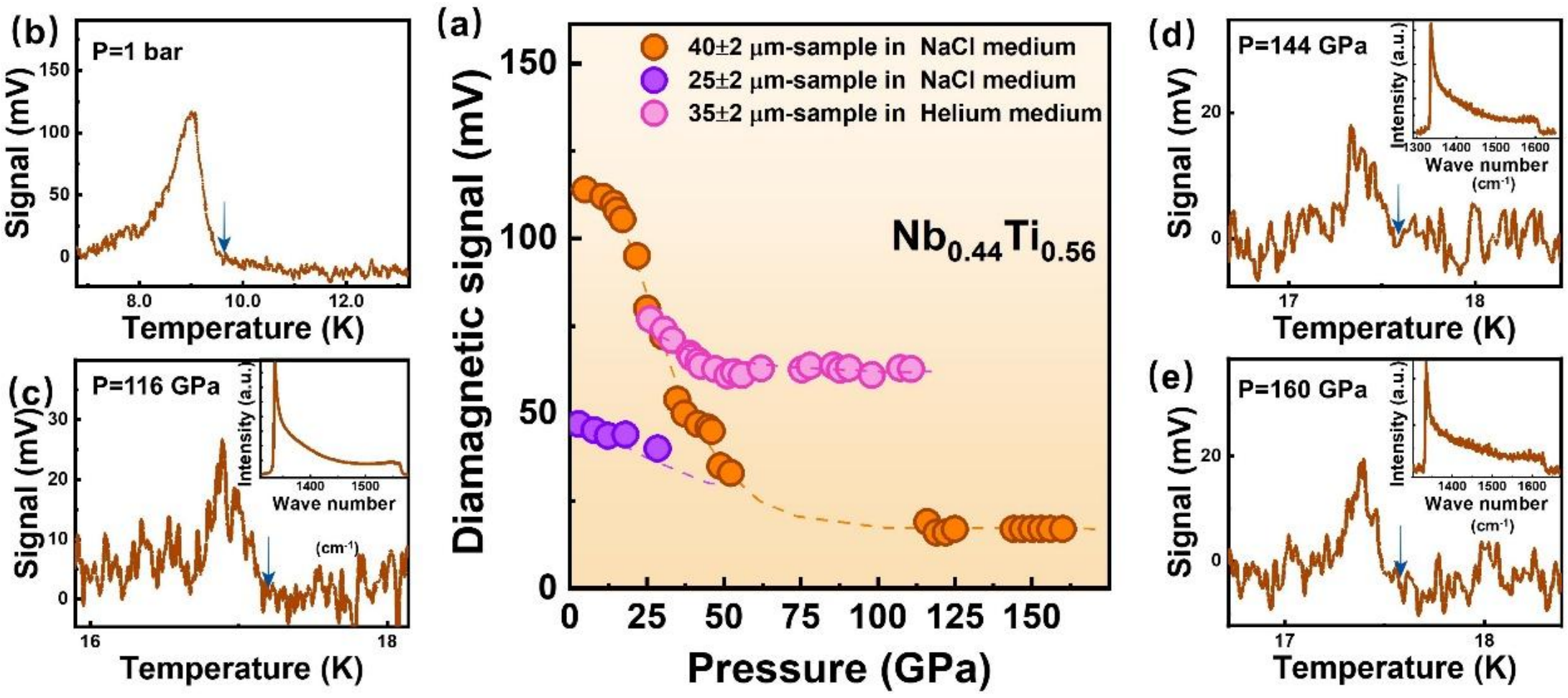

**Figure 3 The diamagnetic signal measured from the compressed superconducting $Nb_{0.44}Ti_{0.56}$ alloy in different pressure transmitting media.** (a) The signal intensity versus pressure obtained from the 40±2 μm sample compressed with 80 μm culets in a NaCl pressure medium (orange solids), 25±2 μm sample compressed with 50 μm culets in a NaCl pressure medium (purple solids), and 35±2 μm sample compressed with 130 μm culets in Helium medium (pink solids). (b)-(e) representative results of the superconducting transition observed at different pressures. The inset of (c)-(e) shows the results of diamond Raman spectroscopy, by which the sample pressure is determined[38, 39].

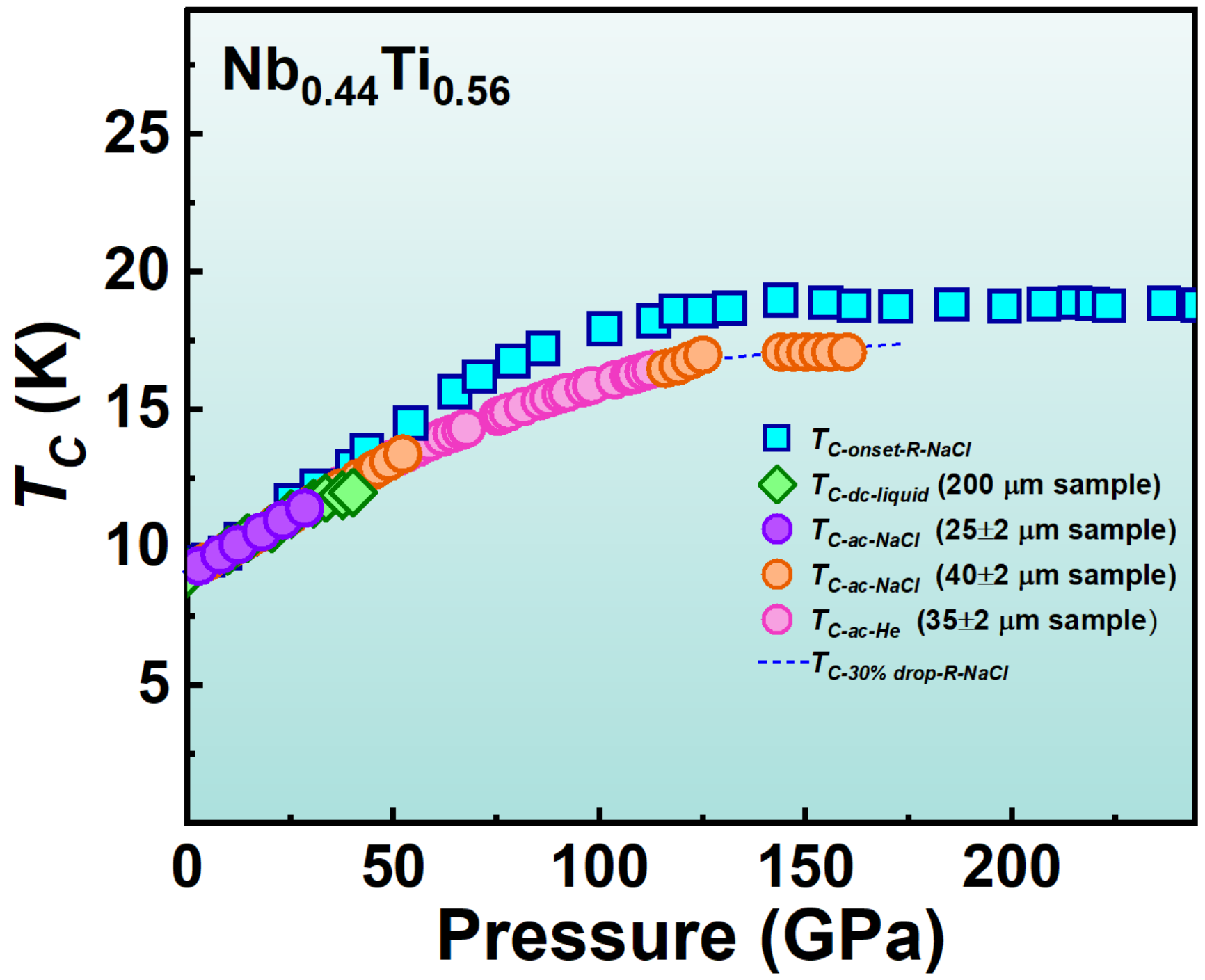

**Figure 4 Pressure dependent superconducting transition temperature ($T_c$) detected by modulated *ac* and *dc* susceptibility as well as resistance measurements for $Nb_{0.44}Ti_{0.56}$.** $T_{c\text{-}ac\text{-}NaCl}$ and $T_{c\text{-}dc\text{-}liquid}$ represent the $T_c$ values obtained from the modulated *ac* and *dc* susceptibility measurements with a NaCl and liquid pressure transmitting media, while $T_{c\text{-}R\text{-}NaCl}$ stands for the onset temperature of superconducting transition determined by resistance measurements with a NaCl pressure transmitting medium. The solids filled with orange and purple are the *ac* susceptibility data of the sample surrounded by a NaCl pressure medium under pressure. The pink solid circles are the *ac* susceptibility data of the sample surrounded by Helium under pressure. The green diamond symbols are the data from the *dc* susceptibility measurements with liquid pressure transmitting medium, while the blue squares are from the resistance measurements. The dashline represents the temperature of 30% resistance drop. The superconducting transition temperature detected by *dc* susceptibility measurements on the compressed $Nb_{0.44}Ti_{0.56}$ sample can be found in SI.

**Table 1** A summary of the modulated *ac* susceptibility results measured from $Nb_{0.44}Ti_{0.56}$, including applied pressure (GPa), culet of the anvils, size/thickness/volume of the sample (μm), $T_c$ (K), signal intensity (SI), signal to noise ratio (SNR) and pressure transmitting medium (PTM).

| Pressure (GPa) | Culet (μm) | Sample size (μm) | Thickness (μm) | Volume (×10³ μm³) | $T_C$ (K) | SI (mV) | SNR | PTM |
|---|---|---|---|---|---|---|---|---|
| 5 | 80 | 40±2 | 8 | 10.0 | 9.5 | 114 | 12.6 | NaCl |
| 11 | 80 | 40±2 | 8 | 10.0 | 10 | 112 | 12.4 | NaCl |
| 13 | 80 | 40±2 | 8 | 10.0 | 10.2 | 110 | 12.2 | NaCl |
| 15 | 80 | 40±2 | 8 | 10.0 | 10.3 | 108 | 12 | NaCl |
| 17 | 80 | 40±2 | 8 | 10.0 | 10.5 | 105 | 11.5 | NaCl |
| 22 | 80 | 40±2 | 8 | 10.0 | 10.9 | 95 | 10.5 | NaCl |
| 25 | 80 | 40±2 | 8 | 10.0 | 11.5 | 80 | 8.9 | NaCl |
| 30 | 80 | 40±2 | 8 | 10.0 | 11.6 | 72 | 8.0 | NaCl |
| 35 | 80 | 40±2 | 8 | 10.0 | 12.0 | 54 | 6.0 | NaCl |
| 37 | 80 | 40±2 | 8 | 10.0 | 12.2 | 50 | 5.3 | NaCl |
| 42 | 80 | 40±2 | 8 | 10.0 | 12.5 | 47 | 5.2 | NaCl |
| 45 | 80 | 40±2 | 8 | 10.0 | 12.8 | 46 | 5.1 | NaCl |
| 47 | 80 | 40±2 | 8 | 10.0 | 12.9 | 45 | 5 | NaCl |
| 49 | 80 | 40±2 | 8 | 10.0 | 13.1 | 35 | 3.9 | NaCl |
| 52 | 80 | 40±2 | 8 | 10.0 | 13.4 | 33 | 3.6 | NaCl |
| 116 | 80 | 40±2 | 8 | 10.0 | 16.5 | 19 | 2.1 | NaCl |
| 119 | 80 | 40±2 | 8 | 10.0 | 16.6 | 16 | 1.8 | NaCl |
| 122 | 80 | 40±2 | 8 | 10.0 | 16.8 | 16 | 1.8 | NaCl |
| 125 | 80 | 40±2 | 8 | 10.0 | 17.0 | 17 | 1.9 | NaCl |
| 144 | 80 | 40±2 | 8 | 10.0 | 17.1 | 17 | 1.9 | NaCl |
| 147 | 80 | 40±2 | 8 | 10.0 | 17.1 | 17 | 1.9 | NaCl |
| 150 | 80 | 40±2 | 8 | 10.0 | 17.5 | 17 | 1.9 | NaCl |
| 153 | 80 | 40±2 | 8 | 10.0 | 17.5 | 17 | 1.9 | NaCl |
| 156 | 80 | 40±2 | 8 | 10.0 | 17.5 | 17 | 1.9 | NaCl |
| 160 | 80 | 40±2 | 8 | 10.0 | 17.5 | 17 | 1.9 | NaCl |
| 3 | 50 | 25±2 | 8 | 3.9 | 9.3 | 51 | 7.3 | NaCl |
| 8 | 50 | 25±2 | 8 | 3.9 | 9.8 | 46 | 6.6 | NaCl |
| 12 | 50 | 25±2 | 8 | 3.9 | 10.1 | 44 | 6.3 | NaCl |
| 18 | 50 | 25±2 | 8 | 3.9 | 10.6 | 42 | 6.0 | NaCl |
| 23 | 50 | 25±2 | 8 | 3.9 | 11 | 38 | 5.4 | NaCl |
| 28 | 50 | 25±2 | 8 | 3.9 | 11.4 | 34 | 4.9 | NaCl |
| 26 | 130 | 35±2 | 8 | 7.7 | 11.4 | 77 | 8.6 | Helium |
| 30 | 130 | 35±2 | 8 | 7.7 | 11.7 | 74 | 8.2 | Helium |
| 33 | 130 | 35±2 | 8 | 7.7 | 11.9 | 71 | 7.9 | Helium |
| 38 | 130 | 35±2 | 8 | 7.7 | 12.4 | 67 | 7.5 | Helium |
| 40 | 130 | 35±2 | 8 | 7.7 | 12.5 | 66 | 7.3 | Helium |
| 42 | 130 | 35±2 | 8 | 7.7 | 12.5 | 65 | 7.2 | Helium |
| 43 | 130 | 35±2 | 8 | 7.7 | 12.6 | 63 | 7.1 | Helium |
| 47 | 130 | 35±2 | 8 | 7.7 | 13.0 | 62 | 7.0 | Helium |
| 51 | 130 | 35±2 | 8 | 7.7 | 13.3 | 60 | 6.8 | Helium |
| 53 | 130 | 35±2 | 8 | 7.7 | 13.5 | 62 | 6.9 | Helium |
| 55 | 130 | 35±2 | 8 | 7.7 | 13.6 | 61 | 6.8 | Helium |
| 62 | 130 | 35±2 | 8 | 7.7 | 14.0 | 63 | 7.0 | Helium |
| 75 | 130 | 35±2 | 8 | 7.7 | 14.7 | 62 | 7.0 | Helium |
| 79 | 130 | 35±2 | 8 | 7.7 | 14.9 | 63 | 7.1 | Helium |
| 85 | 130 | 35±2 | 8 | 7.7 | 15.3 | 64 | 7.1 | Helium |
| 88 | 130 | 35±2 | 8 | 7.7 | 15.4 | 63 | 6.9 | Helium |
| 91 | 130 | 35±2 | 8 | 7.7 | 15.6 | 63 | 7.0 | Helium |
| 98 | 130 | 35±2 | 8 | 7.7 | 15.9 | 61 | 6.8 | Helium |
| 107 | 130 | 35±2 | 8 | 7.7 | 16.2 | 63 | 7.0 | Helium |
| 110 | 130 | 35±2 | 8 | 7.7 | 16.4 | 63 | 7.0 | Helium |